\documentclass{tlp}

\usepackage[english]{babel}
\usepackage{fancyvrb}
\usepackage{tikz}
\usetikzlibrary{patterns}
\usepackage{titletoc}
\usepackage{url}
\usepackage{booktabs}
\usepackage{float}

\usepackage{amsmath}
\usepackage{amssymb}

\newenvironment{example}
{
  \par\vspace{\baselineskip}%
  \textbf{Example}~%
}%
{\par\vspace{\baselineskip}}

\usepackage{color}

\newcommand{\BspPoint}[1]{\filldraw #1 circle(0.04cm)}
\newcommand{\TargetPoint}[1]{\draw #1 circle(0.04cm)}

\newtheorem{Theorem}{Theorem}

\DeclareMathOperator{\qry}{\mathit{?-}}
\DeclareMathOperator{\qryP}{\qry\limits_{p}}

\newcommand{\true}{\mathit{true}}

\newcommand{\authcount}[1]{} 

\usepackage{framed}
{\endMakeFramed}

\begin{document}

\lefttitle{Vandenbroucke and Schrijvers}

\jnlPage{1}{8}
\jnlDoiYr{2022}
\doival{10.1017/xxxxx}

\title[Disjunctive Delimited Control]{Disjunctive Delimited Control\footnote{This article is an extended version of the paper with the same name that first appeared in the LOPSTR 2021 post-proceedings edited
by Emanuele De Angelis and Wim Vanhoof.}}

\begin{authgrp}
\author{\sn{Alexander} \gn{Vandenbroucke}}
\affiliation{Standard Chartered}
\author{\sn{Tom} \gn{Schrijvers}}
\affiliation{KU Leuven}
\end{authgrp}


\history{\sub{xx xx xxxx;} \rev{xx xx xxxx;} \acc{xx xx xxxx}}

\label{firstpage}

\maketitle

\begin{abstract}
Delimited control is a powerful mechanism for programming language extension
which has been recently proposed for Prolog (and implemented in SWI-Prolog).
By manipulating the control flow of a program from inside the language, it
enables the implementation of powerful features, such as tabling, without
modifying the internals of the Prolog engine.
However, its current formulation is inadequate: it does not
capture Prolog's unique non-deterministic nature which allows multiple ways to
satisfy a goal.

This paper fully embraces Prolog's non-determinism with a novel interface for
\emph{disjunctive} delimited control, which gives the programmer not only control over
the sequential (conjunctive) control flow, but also over the non-deterministic
control flow. 
We provide a meta-interpreter that conservatively extends Prolog with delimited
control and show that it enables a range of typical Prolog features and
extensions, now at the library level: findall, cut, branch-and-bound
optimisation, probabilistic programming, \ldots

This paper is under consideration for publication in Theory and Practice of Logic Programming (TPLP).
\end{abstract}


\section{Introduction}

Delimited control is a powerful programming language mechanism for control flow
manipulation that was developed in the late '80s in the context of functional
programming~(\cite{Felleisen:1988,abstracting_control}). \cite{iclp2013} have recently ported this mechanism to Prolog.

Compared to both low-level abstract machine extensions and high-level global
program transformations, delimited control is much more light-weight and robust
for implementing new control-flow and dataflow features.  Indeed, the Prolog
port has enabled powerful applications in Prolog, such as high-level
implementations of both tabling~(\cite{iclp2015}) and algebraic effects \&
handlers~(\cite{iclp2016}). Yet, at the same time, there is much
untapped potential, as the port fails to recognise the unique nature of Prolog when
compared to functional and imperative languages that have previously adopted
delimited control.

Indeed, computations in other languages have only one \emph{continuation},
i.e., one way to proceed from the current point to a result. In contrast, at
any point in a Prolog continuation, there may be multiple ways to proceed and
obtain a result. More specifically, we can distinguish 1) the success or
\emph{conjunctive} continuation which proceeds with the current state of the
continuation; and 2) the failure or \emph{disjunctive} continuation which
bundles the alternative ways to proceed, e.g., if the conjunctive continuation
fails.

The original delimited control only accounts for one continuation, which
Schrijvers et al. have unified with Prolog's conjunctive continuation. More
specifically, for a given subcomputation, they allow to wrest the current
conjunctive continuation from its track, and to resume it at leisure,
however many times as desired.
Yet, this entirely ignores the disjunctive continuation, which remains
as and where it is.

In this work, we adapt delimited control to embrace the whole of Prolog and
capture both the conjunctive and the disjunctive continuations. This makes it
possible to manipulate Prolog's built-in search for custom search strategies
and enables clean implementations of, e.g., \texttt{findall/3} and
branch-and-bound.
This new version of delimited control has an executable specification in the
form of a meta-interpreter (Section~\ref{sec:meta-interpreter}), that can
run both the above examples, amongst others. Appendices to this paper are
available in the extended version~(\cite{DBLP:journals/corr/abs-2108-02972})
and the paper's code is available in the online repository at 
\url{https://github.com/alexandervandenbroucke/tplp-disjunctive-delimited-continuations}.

\section{Overview and Motivation}

We briefly review conjunctive delimited control, explain its obliviousness to 
Prolog disjunctions, and introduce disjunctive delimited control by example.

\subsection{Background: Conjunctive Delimited Control}
\label{sec:bg-conj}

In earlier work, \cite{iclp2013} have introduced a
Prolog-compatible interface for delimited control that consists of two
predicates: \texttt{reset/3} and \texttt{shift/1}.  

\paragraph{Motivation}

While library developers and advanced users typically do not build in new language
features in Prolog, they have traditionally been able to add  
various language extensions by means of Prolog’s rich meta-programming and program
transformation facilities. Examples are definite clause grammars (DCGs),
extended DCGs (\cite{edcgs}), Ciao Prolog’s structured state threading
(\cite{structured_state}) and logical loops (\cite{logical_loops}). However, there
are several important disadvantages to non-local program transformations for
defining new language features: A transformation that combines features can be
quite complex and is fragile under language evolution. Moreover, existing code
bases typically need pervasive changes to, e.g., include DCGs.

Delimited continuations enable new language features at the program level
rather than as program transformations. This makes features based on delimited
continuations more light-weight and more robust with respect to changes, and it
does not require pervasive changes to existing code.

\paragraph{Behavior}

The predicate \texttt{reset(Goal,ShiftTerm,Cont)} executes
\texttt{Goal}, and,
(a) if \texttt{Goal} fails, \texttt{reset/3} also fails;
(b) if \texttt{Goal} succeeds, then \texttt{reset/3} also succeeds and
    unifies \texttt{Cont} and \texttt{ShiftTerm} with \texttt{0};
(c) if \texttt{Goal} calls \texttt{shift(Term)}, then the execution of
    \texttt{Goal} is suspended and \texttt{reset/3} succeeds immediately, 
    unifying \texttt{ShiftTerm} with \texttt{Term} and \texttt{Cont} with the
    remainder of \texttt{Goal}.
The \texttt{shift/reset} pair resembles the more familiar
\texttt{catch/throw} predicates, with the following differences:
\texttt{shift/1} does not copy its argument (i.e., it does not refresh the
variables), it does not delete choice points, and 
also communicates the remainder of \texttt{Goal} to \texttt{reset/3}.

\begin{example}
Consider Definite Clause Grammars (DCGs), a language extension to sequentially
access the elements of an implicit list. It is conventionally defined by a
program transformation that requires special syntax to mark DCG clauses \Verb!H --> B!
and to mark non-DCG goals \Verb!{G}!. The delimited control approach
requires neither. It introduces two new predicates: \Verb!c(E)! consumes the
next element \Verb!E! in the implicit list, and \Verb!phrase(G,Lin,Lout)!  runs
goal \Verb!G! with implicit list \Verb!Lin! and returns unconsumed remainder
\Verb!Lout!. 
For instance, the following predicate implements the grammar $(ab)^n$ and returns
$n$.
\begin{Verbatim}[frame=single]
    ab(0).
    ab(N) :- c(a), c(b), ab(M), N is M + 1.

    ?- phrase(ab(N),[a,b,a,b],[]). 
    N = 2.
 \end{Verbatim}

The two DCG primitives are implemented as follows in terms of \Verb!shift/1! and
\Verb!reset/3!.
\begin{Verbatim}[frame=single]
    c(E) :- shift(c(E)).

    phrase(Goal,Lin,Lout) :-
      reset(Goal,Cont,Term), 
      ( Cont == 0 ->
          Lin = Lout
      ; Term = c(E) ->
          Lin = [E|Lmid],
          phrase(Cont,Lmid,Lout) 
      ).
\end{Verbatim}
In words, \Verb!phrase/3! executes the given goal within a \Verb!reset/3!
and analyzes the possible outcomes. If \Verb!Cont == 0!, this means the goal
succeeds without consuming any input. Then the remainder \Verb!Lout! is equal to the
input list \Verb!Lin!. Alternatively, the execution of the goal has been
suspended midway by the invocation of a \Verb!shift/1! because it wants to
consume an element from the implicit list with \texttt{c/1}. In that case,
\Verb!Term! has been instantiated with a request \Verb!c(E)! for an element
\Verb!E!.
This request is satisfied by instantiating \Verb!E! with the first element
of \Verb!Lin!. Finally, the remainder of the suspended goal, \Verb!Cont! (the continuation),
is resumed with the remainder of the list \Verb!Lmid!.
\end{example}

Other examples of language features implemented in terms of delimited
control are co-routines, algebraic effects~(\cite{iclp2016}) and tabling~(\cite{iclp2015}).

\paragraph{Obliviousness to Disjunctions}

This form of delimited control only captures the conjunctive continuation. For
instance \texttt{reset((shift(a),G1),Term,Cont)} captures in \texttt{Cont} goal
\texttt{G1} that appears in conjunction to \texttt{shift(a)}. In a low-level
operational sense this corresponds to delimited control in other (imperative
and functional) languages where the only possible continuation to capture is
the computation that comes sequentially after the shift. 
Thus this approach is
very useful for enabling conventional applications of delimited control in
Prolog.

In functional and imperative languages delimited control can also be
characterised at a more conceptual level as capturing the entire remainder of a
computation. Indeed, in those languages the sequential
continuation coincides with the entire remainder of a computation. Yet, the
existing Prolog approach fails to capture the entire remainder of a goal, as it
only captures the conjunctive continuation and ignores any disjunctions.
This can be illustrated by the \texttt{reset((shift(a),G1;G2),Term,Cont)} which
only captures the conjunctive continuation \texttt{G1} in \texttt{Cont} 
and not the disjunctive continuation \texttt{G2}. In other words, only the
conjunctive part of the goal's remainder is captured.

This is a pity because disjunctions are a key feature of Prolog and many
advanced manipulations of Prolog's control flow involve manipulating those
disjunctions in one way or another.

\subsection{Delimited Continuations with Disjunction}

This paper presents an approach to delimited control for Prolog that is in line
with the conceptual view that the whole remainder of a goal should be captured,
including in particular the disjunctive continuation.

For this purpose we modify the \texttt{reset/3} interface, where
depending on \texttt{Goal},\\
\texttt{reset(Pattern,Goal,Result)} has three possible outcomes:
\begin{enumerate}
\item
If \texttt{Goal} fails, then the \texttt{reset} succeeds and unifies
\texttt{Result} with \texttt{failure}. For instance,
\begin{Verbatim}[frame=single]
  ?- reset(_,fail,Result).
  Result = failure.
\end{Verbatim}
\item
If \texttt{Goal} succeeds, then 
\texttt{Result} is unified with \texttt{success(PatternCopy,}\\
\texttt{DisjCont)} and the \texttt{reset} succeeds.
Here \texttt{DisjCont} is a goal that represents the disjunctive remainder of
\texttt{Goal}. For instance,
\begin{Verbatim}[frame=single]
  ?- reset(X,(X = a; X = b),Result).
  X = a, Result = success(Y,Y = b).
\end{Verbatim}
Observe that, similar to \texttt{findall/3}, the logical variables in
\texttt{DisjCont} have been renamed apart to avoid interference between the
branches of the computation. To be able to identify any variables of interest
after renaming, we provide \texttt{PatternCopy} as a likewise renamed-apart
copy of \texttt{Pattern}. 

If there is no disjunctive remainder, \texttt{DisjCont} will simply be \texttt{fail}.
\item 
If \texttt{Goal} calls \texttt{shift(Term)}, then the \texttt{reset} succeeds and
\texttt{Result} is unified with \texttt{shift(Term,ConjCont,PatternCopy,DisjCont)}.
This contains in addition to the disjunctive continuation also the conjunctive
continuation. The latter is not renamed apart and can share variables
with \texttt{Pattern} and \texttt{Term}. For instance,
\begin{Verbatim}[frame=single]
  ?- reset(X,(shift(t),X = a; X = b),Result).
  Result = shift(t,X = a, Y, Y = b).
\end{Verbatim}
\end{enumerate}
Note that \texttt{reset(P,G,R)} always succeeds if \texttt{R} is unbound and
never leaves choicepoints.

\paragraph{Encoding \texttt{not/1}}
As a small warm-up exercise, we show how to encode \texttt{not/1}.
\begin{Verbatim}[frame=single]
not(Goal) :-
  reset(_,Goal,Result),
  Result = failure.
\end{Verbatim}
This encoding calls \texttt{Goal} through \texttt{reset/3} and checks that the
result is failure; in this case the pattern argument of \texttt{reset/3} is
irrelevant. If the outcome is anything else, \texttt{not(Goal)} clearly fails
as required.

\paragraph{Encoding \texttt{findall/3}}

Section~\ref{sec:case-studies} presents larger applications of disjunctive delimited control,
but our encoding of 
\texttt{findall/3} with already gives an idea
of the expressive power:
\begin{Verbatim}[frame=single]
findall(Pattern,Goal,List) :-
  reset(Pattern,Goal,Result),
  findall_result(Result,Pattern,List).

findall_result(failure,_,[]).
findall_result(success(PatternCopy,DisjCont),Pattern,List) :-
  List = [Pattern|Tail],
  findall(PatternCopy,DisjCont,Tail).
\end{Verbatim}
This encoding is structured around a \texttt{reset/3} call of the given
\texttt{Goal} followed by a case analysis of the result. Here we assume that
\texttt{shift/1} is not called in \texttt{Goal}, which is a reasonable
assumption for plain \texttt{findall/3}.

\paragraph{Encoding \texttt{!/0}}
Our encoding of the \texttt{!/0} operator illustrates the use of
\texttt{shift/1}:
\begin{Verbatim}[frame=single]
cut :- shift(cut).  

scope(Goal) :-
  copy_term(Goal,Copy),
  reset(Copy,Copy,Result),
  scope_result(Result,Goal,Copy).

scope_result(failure,_,_) :- 
  fail.
scope_result(success(DisjCopy,DisjGoal),Goal,Copy) :- 
  Goal = Copy.
scope_result(success(DisjCopy,DisjGoal),Goal,Copy) :- 
  DisjCopy = Goal,
  scope(DisjGoal).
scope_result(shift(cut,ConjGoal,DisjCopy,DisjGoal),Goal,Copy) :- 
  Copy = Goal,
  scope(ConjGoal).
\end{Verbatim}
The encoding provides \texttt{cut/0} as a substitute for \texttt{!/0}. Where
the scope of regular cut is determined lexically, we use \texttt{scope/1} here
to define it dynamically. For instance, we encode
\begin{center}
\begin{tabular}{ccc}
\begin{minipage}{0.47\textwidth}
\begin{Verbatim}[frame=single]
p(X,Y) :- q(X), !, r(Y).              
p(4,2).

\end{Verbatim}
\end{minipage}
&
as
&
\begin{minipage}{0.47\textwidth}
\begin{Verbatim}[frame=single]
p(X,Y) :- scope(p_aux(X,Y)).
p_aux(X,Y) :- q(X), cut, r(Y).           
p_aux(4,2).
\end{Verbatim}
\end{minipage}
\end{tabular}
\end{center}
The logic of cut is captured in the definition of \texttt{scope/1}; all the
\texttt{cut/0} predicate does is request the execution of a cut with
\texttt{shift/1}.

In \texttt{scope/1}, the \texttt{Goal} is copied to avoid instantiation by any
of the branches. The copied goal is executed inside a \texttt{reset/3} with the
copied goal itself as the pattern. The \texttt{scope\_result/3} predicate
handles the result: 
\begin{itemize}
\item \texttt{failure} propagates with \texttt{fail};
\item \texttt{success} creates a disjunction to either unify the initial goal with the now instantiated copy
      to propagate bindings, or to invoke the disjunctive continuation;
\item \texttt{shift(cut)} discards the disjunctive continuation and proceeds
      with the conjunctive continuation only.
\end{itemize}

\paragraph{Encoding Non-Backtrackable State}
Disjunctive delimited control can also be used to express custom dataflows,
such as non-backtrackable state. 
For the sake of simplicity, we encode here a single nameless global state.
This can be easily extended to support SICStus Prolog's blackboard primitives
or SWI-Prolog's non-backtrackable variables.

The state is read and written with respectively \texttt{get/1} and \texttt{put/1}.
For instance, predicate \texttt{q/1} writes 1 to the global state in the first
clause, fails and backtracks to the next clause to read and use the value of the global state.
\begin{Verbatim}[frame=single]
q(_) :- put(1), fail.
q(Y) :- get(X), Y is X + 1.
\end{Verbatim}
The \texttt{run\_state(Goal,Initstate,FinalState)} is the analog of DCG's \texttt{phrase/3}
for running a goal with a given initial value for the global state and resulting final value.
\begin{Verbatim}[frame=single]
?- run_state(q(Y),0,S).
Y = 2,
S = 1.
\end{Verbatim}
As this query illustrates, the value 1 written in the first branch survives the
backtracking and is still available in the second branch.

Figure~\ref{fig:state} shows the implementation of this non-backtrackable state interface.
Both \texttt{get/1} and \texttt{put/1} are defined in terms of \texttt{shift/1}.
The \texttt{run\_state/3} predicate calls the goal inside \texttt{reset/3} and subsequently
handles the result with \texttt{state\_handler/5}. The latter auxiliary predicate
essentially acts as a meta-interpreter for the non-deterministic structure of the goal without resorting to Prolog's 
underlying backtracking. This way it avoids backtracking over the global state.

\begin{figure}
\begin{Verbatim}[frame=single]
get(X) :- shift(get(X)).
put(X) :- shift(put(X)).

run_state(Goal,InitState,FinalState) :-
    copy_term(Goal,Pattern),
    reset(Pattern,Goal,Result),
    state_handler(Goal,Pattern,Result,InitState,FinalState).

state_handler(Goal,Pattern,success(PatternCopy,Branch),SIn,SOut) :-
    ( Goal = Pattern, SIn = SOut
    ; reset(PatternCopy,Branch,Result),
      state_handler(Goal,PatternCopy,Result,SIn,SOut)
    ).

state_handler(Goal,Pattern,shift(put(X),Cont,PatternCopy,Branch),_SIn,SOut) :-
    SMid = X,
    PatternCopy = Pattern,
    reset(Pattern,(Cont ; Branch),NewResult),
    state_handler(Goal,Pattern,NewResult,SMid,SOut).

state_handler(Goal,Pattern,shift(get(X),Cont,PatternCopy,Branch),SIn,SOut) :-
    X = SIn,
    PatternCopy = Pattern,
    reset(Pattern,(Cont ; Branch),NewResult),
    state_handler(Goal,Pattern,NewResult,SIn,SOut).
state_handler(_,_,failure,_,_,_) :- fail.
\end{Verbatim}
\caption{Encoding of non-backtrackable state}\label{fig:state}
\end{figure}

\section{Meta-Interpreter Semantics}
\label{sec:meta-interpreter}

We provide an accessible definition of disjunctive delimited control in
the form of a meta-interpreter.
Broadly speaking, it consists of two parts: the core interpreter, and a top
level predicate to initialise the core and interpret the results.

\subsection{Core Interpreter} 

\begin{figure}[b!]
\begin{Verbatim}[frame=single,numbers=left]
eval([],PatIn,Disj,PatOut,Result) :- !,
    PatOut = PatIn,
    Disj   = alt(BranchPatIn,BranchGoal),
    Result = success(BranchPatIn,BranchGoal).
eval([true|Conj],PatIn,Disj,PatOut,Result) :- !,
    eval(Conj,PatIn,Disj,PatOut,Result).
eval([(G1,G2)|Conj],PatIn,Disj,PatOut,Result) :- !,
    eval([G1,G2|Conj],PatIn,Disj,PatOut,Result).
eval([fail|_Conj],_,Disj,PatOut,Result) :- !,
    backtrack(Disj,PatOut,Result).
eval([(G1;G2)|Conj],PatIn,Disj,PatOut,Result) :- !,
    copy_term(alt(PatIn,conj([G2|Conj])),Branch),
    disjoin(Branch,Disj,NewDisj),
    eval([G1|Conj],PatIn,NewDisj,PatOut,Result).
eval([conj(Cs)|Conj],PatIn,Disj,PatOut,Result) :- !,
    append(Cs,Conj,NewConj),
    eval(NewConj,PatIn,Disj,PatOut,Result).
eval([shift(Term)|Conj],PatIn,Disj,PatOut,Result) :- !,
    PatOut = PatIn,
    Disj   = alt(BranchPatIn,Branch),
    Result = shift(Term,conj(Conj),BranchPatIn,Branch).
eval([reset(RPattern,RGoal,RResult)|Conj],PatIn,Disj,PatOut,Result):- !,
    copy_term(RPattern-RGoal,RPatIn-RGoalCopy),
    empty_alt(RDisj),
    eval([RGoalCopy],RPatIn,RDisj,RPatOut,RResultFresh),
    eval([RPattern=RPatOut,RResult=RResultFresh|Conj]
         ,PatIn,Disj,PatOut,Result).
eval([Call|Conj],PatIn,Disj,PatOut,Result) :- !,
    findall(Call-Body,clause(Call,Body), Clauses),
    ( Clauses = [] -> backtrack(Disj,PatOut,Result)
    ; disjoin_clauses(Call,Clauses,ClausesDisj),
      eval([ClausesDisj|Conj],PatIn,Disj,PatOut,Result)
    ).
\end{Verbatim}
\caption{Meta-Interpreter Core}\label{fig:eval}
\end{figure}

\begin{figure}[t!]
\begin{Verbatim}[frame=single,numbers=left]
backtrack(Disj,PatOut,Result) :-
    ( empty_alt(Disj) ->
       Result = failure
    ; Disj = alt(BranchPatIn,BranchGoal) ->
       empty_alt(EmptyDisj),
       eval([BranchGoal],BranchPatIn,EmptyDisj,PatOut,Result)
    ).

empty_alt(alt(_,fail)).

disjoin(alt(_,fail),Disjunction,Disjunction) :- !.
disjoin(Disjunction,alt(_,fail),Disjunction) :- !.
disjoin(alt(P1,G1),alt(P2,G2),Disjunction) :-
    Disjunction = alt(P3, (P1 = P3, G1 ; P2 = P3, G2)).

disjoin_clauses(_G,[],fail) :- !.
disjoin_clauses(G,[GC-Clause],(G=GC,Clause)) :- !.
disjoin_clauses(G,[GC-Clause|Clauses], ((G=GC,Clause) ; Disj)) :-
    disjoin_clauses(G,Clauses,Disj).
\end{Verbatim}
\caption{Auxiliary Predicates for Meta-Interpreter Core}\label{fig:aux}
\end{figure}

Figure~\ref{fig:eval} defines the interpreter's core predicate,
\texttt{eval(Conj, PatIn, Disj,\\PatOut, Result)}. 
It captures the behaviour of \texttt{reset(Pattern,Goal,Result)} where the goal
is given in the form of a list of goals, \texttt{Conj}, together
with the alternative branches, \texttt{Disj}. The latter is
renamed apart from \texttt{Conj} to avoid conflicting
instantiations.

The pattern that identifies the variables of interest (similar to
\texttt{findall/3}) is present in three forms. Firstly, \texttt{PatIn} is
an input argument that shares the variables of interest with \texttt{Conj} (but
not with \texttt{Disj}). Secondly, \texttt{PatOut} outputs the instantiated
pattern when the goal succeeds or suspends on a \texttt{shift/1}. Thirdly, the
alternative branches \texttt{Disj} are of the form
\texttt{alt(BranchPatIn,BranchGoal)} with their own copy of the pattern.

When the conjunction is empty (1--4), the output pattern is unified with the
input pattern, and \texttt{success/2} is populated with the information from
the alternative branches.

When the first conjunct is \texttt{true/0} (5--6), it is dropped and the meta-interpreter
proceeds with the remainder of the conjunction. When it is a
composite conjunction \texttt{(G1,G2)} (7--8), the individual components are added separately to the
list of conjunctions. 

When the first conjunct is \texttt{fail/0} (9--10), the meta-interpreter
backtracks explicitly by means of auxiliary predicate \texttt{backtrack/3} (see Fig.~\ref{fig:aux}).
If there is no alternative branch, it sets the \texttt{Result} to
\texttt{failure}.

Otherwise, it resumes with the alternative branch.
Note that by managing its own backtracking, \texttt{eval/5} is entirely
deterministic with respect to the meta-level Prolog system.

When the first conjunct is a disjunction \texttt{(G1;G2)} (11--14), the
meta-interpreter adds (a renamed apart copy of) \texttt{(G2,Conj)} to the
alternative branches with \texttt{disjoin/3} (see Fig.~\ref{fig:aux}) and proceeds with
\texttt{[G1|Conj]}.

Note that we have introduced a custom built-in \texttt{conj(Conj)}  that turns
a list of goals into an actual conjunction. It is handled (15--17) by
prepending the goals to the current list of conjuncts, and never actually builds
the explicit conjunction.

When the first goal is \texttt{shift(Term)} (18--21), this is handled similarly
to an empty conjunction, except that the result is a \texttt{shift/4} term
which contains \texttt{Term} and the remainder of the conjunction in addition
the branch information.

When the first goal is a \texttt{reset(RPattern,RGoal,RResult)} (22--27), the
meta-interpreter sets up an isolated call to \texttt{eval/5} for this goal.
When the call returns, the meta-interpreter passes on the results and resumes
the current conjunction \texttt{Conj}. Notice that we are careful
that this does not result in meta-level failure by meta-interpreting the
unification.

Finally, when the first goal is a call to a user-defined predicate (28--33), the
meta-interpreter collects the bodies of the predicate's clauses whose head
unifies with the call. If there are none, it backtracks explicitly. Otherwise,
it builds an explicit disjunction with \texttt{disjoin\_clauses} (see Fig.~\ref{fig:aux}), which it
pushes on the conjunction stack.

An example execution trace of the interpreter can be found in
\cite[Appendix C]{DBLP:journals/corr/abs-2108-02972}.

\subsection{Toplevel} The \texttt{toplevel(Goal)}-predicate (see Fig.~\ref{fig:top}) initialises
the core interpreter with a conjunction containing only the given goal, the
pattern and pattern copy set to (distinct) copies of the goal, and an empty
disjunction.  It interprets the result by non-deterministically producing all
the answers to \texttt{Goal} and signalling an error for any unhandled
\texttt{shift/1}.
\begin{figure}[t]
\begin{Verbatim}[frame=single]
toplevel(Goal) :-
    copy_term(Goal,GoalCopy),
    PatIn = GoalCopy,
    empty_alt(Disj),
    eval([GoalCopy],PatIn,Disj,PatOut,Result),
    ( Result = success(BranchPatIn,Branch) ->
        ( Goal = PatOut ; Goal = BranchPatIn, toplevel(Branch))
    ; Result = shift(_,_,_,_) ->
        write('toplevel: uncaught shift/1.\n'), fail
    ; Result = failure ->
        fail
    ).
\end{Verbatim}
\caption{Meta-Interpreter Toplevel}\label{fig:top}
\end{figure}

\subsection{Performance Discussion}

Our meta-interpreter is an executable specification that allows prototyping new
language features on top of disjunctive control. Yet, it clearly has
scalability issues. Notably, the interpreter makes use of \texttt{copy\_term/2}
at every disjunction. This can easily lead to a quadratic runtime on its own.
Moreover, several applications (like our encoding of cut) add further uses of
\texttt{copy\_term/2} on top of that.

These scalability issues can be partly mitigated by providing native (e.g.,
WAM-level) support for disjunctive control. We expect that more significant
(algorithmic) gains can be obtained by providing native support for new
language features implemented with delimited control, effectively moving them
from the prototyping stage to the production stage. This way unnecessary
overhead, that stems from the generic nature of disjunctive control, can be
removed in favour of exploiting feature-specific properties.
An example of a language feature that has undergone a similar evolution is SWI-Prolog's
tabling~\citep{iclp2015}, which was originally implemented with conjunctive delimited control and
pure Prolog datastructures, and later several of its components were re-implemented
in C for greater performance.

\section{Case Studies}
\label{sec:case-studies}

To illustrate the usefulness and practicality of our approach, we present two
case studies that use the new \texttt{reset/3} and \texttt{shift/1}:
branch-and-bound search and probalistic programming in both the PRISM and ProbLog flavors.

\subsection{Branch-and-Bound: Nearest Neighbour Search}

Branch-and-bound is a well-known general optimisation strategy, where the
solutions in certain areas or branches of the search space are known to be
bounded.
Such branches can be pruned, when their bound does not improve upon a
previously found solution, eliminating large swaths of the search space in a
single stroke.

We provide an implementation\footnote{The code in Figures~\ref{fig:branch-and-bound}
and \ref{fig:nearest-neighbour} uses if-then-else (\texttt{ -> ; })
which is not supported by the meta-interpreter. We use it here to simplify the
presentation, as the code could be easily re-written without if-then-else.}
of branch-and-bound (see Figure~\ref{fig:branch-and-bound}) that is generic,
i.e., it is not specialised for any application.
In particular it is not specific to nearest neighbour search, the problem
on which we demonstrate the branch-and-bound approach here.

\begin{figure}[t]
  \centering
\begin{minipage}{0.95\textwidth}  
\begin{Verbatim}[frame=single]
bound(V) :- shift(V).
  
bb(Value,Data,Goal,Min) :-
    reset(Data,Goal,Result),
    bb_result(Result,Value,Data,Min).

bb_result(success(BranchCopy,Branch),Value,Data,Min) :-
  ( Data @< Value -> bb(Data,BranchCopy,Branch,Min)
  ; bb(Value,BranchCopy,Branch,Min)
  ).
bb_result(shift(ShiftTerm,Cont,BranchCopy,Branch),Value,Data,Min) :-
  (  ShiftTerm @< Value ->
     bb(Value,Data,(Cont ; (BranchCopy = Data,Branch)),Min)
  ;  bb(Value,BranchCopy,Branch,Min)
  ).
bb_result(failure,Value,_Data,Min) :- Value = Min.
\end{Verbatim}
\end{minipage}
\caption{Branch-and-Bound Effect Handler.}
\label{fig:branch-and-bound}
\end{figure}

The framework requires minimal instrumentation: it suffices to begin every
prunable branch with \texttt{bound(V)}, where \texttt{V} is a lower
bound on the values in the branch.\footnote{The framework searches for a
  minimal solution.}
\begin{enumerate}
\item If the \texttt{Goal} succeeds normally (i.e., \texttt{Result} is
  \texttt{success}), then \texttt{Data} contains a new solution, which is only
  accepted if it is an improvement over the existing \texttt{Value}.
  The handler then tries the next \texttt{Branch}.
\item If the \texttt{Goal} calls \texttt{bound(V)}, \texttt{V} is compared
  to the current best \texttt{Value}:
  \begin{itemize}
  \item if it is less than the current value, then \texttt{Cont} could
    produce a solution that improves upon the current value, and thus must
    be explored.
    The alternative \texttt{Branch} is disjoined to \texttt{Cont}, and
    \texttt{DataCopy} is restored to \texttt{Data} (ensuring that a future
    \texttt{reset/3} copies the right variables);
  \item if it is larger than or equal to the current value, then \texttt{Cont}
    can be safely discarded.
  \end{itemize}
\item Finally, if the goal fails entirely, \texttt{Min} is the current
  minimum \texttt{Value}.

\end{enumerate}


\begin{figure}[tb!]
\begin{minipage}{\textwidth}
\begin{Verbatim}[frame=single]
nn((X,Y),BSP,D-(NX,NY)) :-
    ( BSP = xsplit((SX,SY),Left,Right) ->
        DX is X - SX, 
        branch((X,Y), (SX,SY), Left, Right, DX, D-(NX,NY))
    ; BSP = ysplit((SX,SY),Up,Down) ->
        DY is Y - SY, 
        branch((X,Y), (SX,SY), Up, Down, DY, D-(NX,NY))
    ).

branch((X,Y), (SX,SY), BSP1, BSP2, D, Dist-(NX,NY)) :-
    ( D < 0 -> % Find out which partition contains (X,Y).
        TargetPart = BSP1, OtherPart = BSP2, BoundaryDistance is -D
    ;  
        TargetPart = BSP2, OtherPart = BSP1, BoundaryDistance is D
    ),
    ( nn((X,Y), TargetPart, Dist-(NX,NY))
    ; Dist is (X - SX) * (X - SX) + (Y - SY) * (Y - SY),
      (NX,NY) = (SX,SY)
    ; bound(BoundaryDistance-nil),
      nn((X,Y), OtherPart,Dist-(NX,NY))
    ).

run_nn((X0,Y0),BSP,(NX,NY)) :-
    toplevel(bb(10-nil,D-(X,Y),nn((X0,Y0),BSP,D-(X,Y)),_-(NX,NY))).
\end{Verbatim}
\end{minipage}
\caption{2D Nearest Neighbour Search with Branch-and-Bound.}
\label{fig:nearest-neighbour}
\end{figure}
\paragraph{Nearest Neighbour Search} The code in
Figure~\ref{fig:nearest-neighbour} shows how the branch and bound
framework efficiently solves the problem of finding the point (in a given
set) that is nearest to a given target point on the Euclidean plane.

The \texttt{run\_nn/3} predicate takes a point \texttt{(X,Y)},
a Binary Space Partitioning (BSP)-tree\footnote{A BSP-tree is a tree that
  recursively partitions a set of points on the Euclidean plane, by picking
  points and alternately splitting the plane along the x- or y-coordinate of
  those points. Splitting along the x-coordinate produces an \texttt{xsplit/3}
  node, a split along the y-coordinate produces a \texttt{ysplit/3} node.}
that represents the set of points, and returns the point, nearest to
\texttt{(X,Y)}.
The algorithm implemented by \texttt{nn/3} recursively descends the BSP-tree.
At each node it first tries the partition to which the target point belongs,
then the point in the node, and finally the other partition.
For this final step we can give an easy lower bound: any point in the
other partition must be at least as far away as the (perpendicular) distance
from the given point to the partition boundary.

\begin{figure}[tb!]
  \centering
  \begin{tikzpicture}[scale=2.2]
    \draw[pattern=north west lines, pattern color=black,opacity=0.5]
      (-1,-1) -- (0,-1) -- (0,1) -- (-1,1) -- cycle;
    \draw (-1,-1) -- (1,-1) -- (1,1) -- (-1,1) -- cycle;
    \TargetPoint{(1,0.1)}   node[anchor=east]  {(1,0.1)};
    \BspPoint{(0,0)}        node[anchor=west]  {(0,0)};
    \BspPoint{(0.5,0.5)}    node[anchor=south] {(0.5,0.5)};
    \BspPoint{(-0.5,0)}     node[anchor=south] {(-0.5,0.5)};
    \BspPoint{(-0.75,-0.5)} node[anchor=west]  {(-0.75,-0.5)};
    \draw (0,-1) -- (0,1);
    \draw (0,0.5) -- (1,0.5);
    \draw (-1,0) -- (0,0);
    \draw (-0.75,0) -- (-0.75,-1);
  \end{tikzpicture}
  \caption{Nearest-Neighbour Search using a BSP-tree}
  \label{fig:bsp}
\end{figure}

As an example, we search for the point nearest to $(1,0.1)$ in the set
$\{(0.5,0.5),$ $(0,0),$ $(-0.5,0),$ $(-0.75,-0.5)\}$.
Figure~\ref{fig:bsp} shows a BSP-tree containing these points,
the solid lines demarcate the partitions.
The algorithm visits the points $(0.5,0.5)$ and $(0,0)$, in that order.
The shaded area is never visited, since the distance from (1,0.1) to the
vertical boundary through $(0,0)$ is greater than the distance to $(0.5,0.5)$
(1 and about 0.64).
The corresponding call to \texttt{run\_nn/3} is:
\begin{Verbatim}[frame=single]
  ?- BSP = xsplit((0,0),
            ysplit((-0.5,0),leaf,xsplit((-0.75,-0.5),leaf,leaf)),
            ysplit((0.5,0.5),leaf,leaf)),
     run_nn((1,0.1),BSP,(NX,NY)).
  NX = NY, NY = 0.5.
\end{Verbatim}

\subsection{Probabilistic Programming}
Probabilistic programming languages (PPLs) are programming languages 
designed for probabilistic modelling.
In a probabilistic model, components behave in a variety of ways---just like in a
non-deterministic model---but do so with a certain probability.

Instead of a single deterministic value, the execution of a probabilistic
program results in a probability distribution of a set of values.
This result is produced by probabilistic
\emph{inference}~(\cite{DBLP:conf/aistats/WoodMM14}),
for which there are many strategies
and algorithms, the discussion of which is out of scope here.
Instead, we focus on two concrete probabilistic \emph{logic} programming
languages: PRISM~(\cite{DBLP:conf/iclp/Sato09}) and 
PRISM (\cite{DBLP:journals/tplp/FierensBRSGTJR15}).

\paragraph{PRISM-Style Probabilistic Logic Programming}
A PRISM program 
looks just like a regular Prolog program
extended with two special predicates:
\begin{itemize}
\paragraph{PRISM-Style Probabilistic Logic Programming}
\item
  \texttt{values\_x(Switch,Values,Probabilities)} This predicate defines a
  probabilistic switch \texttt{Switch}, that can assume a value from
  \texttt{Values} with the probability that is given at the corresponding
  position in \texttt{Probabilities} (the contents of \texttt{Probabilities}
  should sum to one).
\item \texttt{msw(Switch,Value)} This predicate samples a value
  \texttt{Value} from a switch \texttt{Switch}.
  For instance, if the program contains a switch declared as
  \texttt{values\_x( coin, [h,t], [0.4,0.6])}, then \texttt{msw(coin,V)}
  assigns \texttt{h} (for heads) to \texttt{V} with probability 0.4, and
  \texttt{t} (for tails) with probability 0.6.
  Remark that each distinct call to \texttt{msw} leads to a different sample
  from that switch.
  For instance, in the query \texttt{msw(coin,X),msw(coin,Y)}, the outcome
  could be either \texttt{(X = h, Y = h)},\texttt{(X = t, Y = t)}, \texttt{(X = h, Y = t)} or
  \texttt{(X = t,Y = h)}.
\end{itemize}
Consider the following PRISM program, the running example for this section:
\begin{Verbatim}[frame=single]
  values_x(coin1,[h,t],[0.5,0.5]).
  values_x(coin2,[h,t],[0.4,0.6]).
  twoheads :- msw(coin1,h),msw(coin2,h).
  onehead :- msw(coin1,V), (V = t, msw(coin2,h) ; V = h).
\end{Verbatim}
This example defines two predicates: \texttt{twoheads} which is true if both
coins are heads, and \texttt{onehead} which is true if either coin is heads.
However, note the special structure of \texttt{onehead}: PRISM requires the
\emph{exclusiveness condition}, that is, branches of a disjunction cannot be
both satisfied at the same time.
The simpler goal \texttt{msw(coin1,heads) ; msw(coin2, heads)} violates this
assumption.

\begin{figure}[tb!]
\begin{Verbatim}[frame=single,numbers=left]
msw(Key,Value) :- shift(msw(Key,Value)).

prism(Goal) :-
    prob(Goal,ProbOut),
    write(Goal), write(': '), write(ProbOut), write('\n').

prob(Goal,ProbOut) :-
    copy_term(Goal,GoalCopy),
    reset(GoalCopy,GoalCopy,Result),
    analyze_prob(GoalCopy,Result,ProbOut).

analyze_prob(_,failure,0.0).
analyze_prob(_,success(_,_),1.0).
analyze_prob(_,shift(msw(K,V),C,_,Branch),ProbOut) :-
    values_x(K,Values,Probabilities),
    msw_prob(V,C,Values,Probabilities,0.0,ProbOfMsw),
    prob(Branch,BranchProb),
    ProbOut is ProbOfMsw + BranchProb.
\end{Verbatim}
\caption{An implementation of PRISM-style probabilistic logic programming.}
\label{fig:prob-prog}
\end{figure}

The code in Figure~\ref{fig:prob-prog} interprets this program.
Line 1 defines \texttt{msw/2} as a simple shift. Next, lines 3--5 define 
the \texttt{prism/1} wrapper predicate that computes and prints a goal's
probability. 
Lines 7--10 install a \texttt{reset/3} call over the goal, and analyse the
result.
The result is analysed in the remaining lines:
A \emph{failure} never succeeds, and thus has success probability 0.0 (line 12).
Conversely, a successful computation has a success probability of 1.0 (line 13).
Finally, the probability of a switch (lines 14-18) is the sum of the
probability of the remainder of the program given each possible value of the
switch multiplied with the probability of that value, and summed with the
probability of the alternative branch.

The predicate \texttt{msw\_prob} finds the joint probability of all choices.
It iterates over the list of values, and sums the probability of their
continuations.
\begin{Verbatim}[frame=single]
  msw_prob(_,_,[],[],Acc,Acc).
  msw_prob(V,C,[Value|Values],[Prob|Probs],Acc,ProbOfMsw) :-
    prob((V = Value,C),ProbOut),
    msw_prob(V,C,Values,Probs,Prob*ProbOut + Acc,ProbOfMsw).
\end{Verbatim}

Now, we can compute the probabilities of the two predicates above:
\begin{Verbatim}[frame=single]
  ?- toplevel(prism(twoheads)).
  twoheads: 0.25
  ?- toplevel(prism(onehead)).
  onehead: 0.75
\end{Verbatim}

\paragraph{ProbLog-Style Probabilistic Logic Programming}
We now encode the loop-free, definite fragment of ProbLog on top of the above
encoding of PRISM. Our encoding uses a different syntax for probabilistic facts
than ProbLog:\footnote{This could be hidden with syntactic sugar based on term expansion.}
\begin{Verbatim}[frame=single]
  % Original ProbLog           % Encoding
  0.5 :: heads1.               values_x(heads1,[t,f],[0.5,0.5]).         
  ?- heads1.                   ?- fact(heads1).
\end{Verbatim}
For the declaration of probabilistic facts we use the PRISM notation (because that
is what we are leveraging underneath). For the invocation of these facts, we use a 
special \texttt{fact/1} predicate.

Semantically, ProbLog distinguishes itself on several accounts from PRISM.
Consider the following variant of \texttt{twoheads/0}.
\begin{Verbatim}[frame=single]
  twoheads1 :- fact(heads1), fact(heads1).
\end{Verbatim}
PRISM assigns the probability 0.25 to \texttt{twoheads1} because it treats the
two occurrences of \texttt{heads1} as independent samples. In contrast,
ProbLog treats them as referring to the same sample and thus assigns
probability 0.5 to \texttt{twoheads1}.

ProbLog also does not require the branches of disjunctions to be mutually
exclusive. Consider the following variant of \texttt{onehead/0}.
\begin{Verbatim}[frame=single]
  onehead1 :- fact(heads1); fact(heads1).
\end{Verbatim}
In PRISM, \texttt{onehead1} is not well-defined because the two branches are
not exclusive. ProbLog in contrast considers \texttt{onehead1} to be true when
\texttt{heads1} is true, which has probability 50\%.  The second branch does
not affect the probability; it is redundant.

\begin{figure}[tb!]
\begin{Verbatim}[frame=single]
problog(Goal) :- 
    problog(Goal,[]).

problog(Goal,Pc) :-
    reset(Goal,Goal,Result),
    analyze_problog(Result,Pc).

analyze_problog(success(_,_),_Pc).
analyze_problog(failure,_Pc) :- 
    fail.
analyze_problog(shift(fact(F,V),C,_,Branch),Pc) :-
    member(F-V,Pc),
    problog((C;Branch),Pc).
analyze_problog(shift(fact(F,V),C,_,Branch),Pc) :-
    not(member(F-_,Pc)),
    msw(F,V),
    problog((C;Branch),[F-V|Pc]).

fact(F) :- 
    shift(fact(F,V)), 
    V = t.
\end{Verbatim}
\caption{An implementation of PRISM-style probabilistic logic programming.}
\label{fig:problog}
\end{figure}

We implement this ProbLog semantics in the \texttt{problog/1} predicate as a
``pre-processor'' for the \texttt{prism/1} encoding of PRISM. Hence, a 
toplevel ProbLog goals \texttt{G} is meant to be called as \texttt{prism(problog(G))}.

The main work of \texttt{problog/1} is done by \texttt{problog/2} which keeps track of a list
\texttt{Pc} of \texttt{F-V} pairs of already sampled probabilistic facts
\texttt{F} and their sampled value \texttt{V}; initially this list is empty. 
The \texttt{problog/2} predicate calls the current goal with \texttt{reset/3}
and analyzes the result. In case of success or failure, it propagates that
success or failure, which means it will be handled by \texttt{prism/1}.

In case the goal calls a probabilistic fact with \texttt{fact/1}, this results
in a \texttt{shift/1} which is intercepted by \texttt{reset/3} and handled in
one of two ways. If the fact was already sampled before and its value is thus
available in the \texttt{Pc} list, the computation proceeds with that value.
Otherwise, the \texttt{msw/2} predicate is used to sample
the fact's value, which is stored in the \texttt{Pc} list for future uses, and
the computation proceeds accordingly. The treatment of \texttt{msw/2} and 
working out the probabilities is delegated to \texttt{prism/1}.

The interplay between \texttt{problog/1} and \texttt{prism/1} is somewhat
subtle. So let us consider what happens in the case of the two example queries
above.

\begin{itemize}
\item \texttt{fact(heads1), fact(heads1)}
  The first occurrence of \texttt{heads1} appeals to \texttt{msw(heasd1,V)},
  but the second only observes the already sampled value. Hence, \texttt{prism}
  only sees one sampling and assigns probability 0.5.
\begin{Verbatim}[frame=single]
    ?- solutions(prism(problog(twoheads1))).
    problog(twoheads1): 0.5
\end{Verbatim}

\item \texttt{fact(heads1); fact(heads1)}
  The first occurrence of \texttt{heads1} appeals to \texttt{msw(heasd1,V)} 
  and essentially rearranges the goal to
\begin{Verbatim}[frame=single]
    ?- prism(
         msw(heads1,V),
         problog((V = t ; fact(heads1)), [heads1-V])
       ).
\end{Verbatim}
  When the sampling yields value \texttt{t}, the \texttt{V = t} unification in the left branch succeeds and the alternative
  branch is discarded. When the sampling yields value \texttt{f}, the unification fails and the right branch is executed:
  The second and remaining occurrence of \texttt{fact(heads1)} now consults the recorded \texttt{f} value---rather than sampling again---and
  also fails because it is not \texttt{t}. Hence, overall there is one success and this success involved one sampling with probability 0.5.
\begin{Verbatim}[frame=single]
    ?- solutions(prism(problog(onehead1))).
    problog(onehead1): 0.5
\end{Verbatim}
\end{itemize}


%

\section{Properties of the Meta-Interpreter}
In this section we establish two important correctness properties of our
meta-interpreter with respect to standard SLD resolution. Together these
establish that disjunctive delimited control is a conservative extension. This
means that programs that do not use the feature behave the same as before.

The proofs of these properties are in \cite[Appendix A]{DBLP:journals/corr/abs-2108-02972}.
The first theorem establishes the soundness of the meta-interpreter, i.e.,
if a program (not containing \texttt{shift/1} or \texttt{reset/3}) evaluates
to success, then an SLD-derivation of the same answer must exist.

\begin{Theorem}[Soundness]
  For all lists of goals $[A_1,\ldots,A_n]$,
  terms $\alpha,\beta,\gamma,\nu$, variables $P,R$
  conjunctions $B_1,\ldots,B_m$; $C_1,\ldots,C_k$ and substitutions $\theta$,
  if
  \[
    \begin{array}{l}
      ?- \mathit{eval}([A_1,\ldots,A_n],\alpha,\mathit{alt}(\beta,(B_1,\ldots,B_m)),P,R).\\
      P = \nu, R = \mathit{success(\gamma,C_1,\ldots,C_k)}.
    \end{array}
  \]
  and the program contains neither \texttt{shift/1} nor \texttt{reset/3},
  then SLD-resolution\footnote{Standard SLD-resolution, augmented with
    disjunctions and \texttt{conj/1} goals.}
    finds the following derivation:
  \[
    \begin{array}{c}
      \leftarrow (A_1,\ldots,A_n,\true) ; (\alpha = \beta,B_1,\ldots,B_m)\\
      \vdots\\
      \square\\
      \text{(with solution $\theta$ s.t. $\alpha\theta = \nu$)}
    \end{array}
  \]
\end{Theorem}
Conversely, we want to argue that the meta-interpreter is complete,
i.e., if SLD-derivation finds a refutation, then
meta-interpretation---provided that it terminates---must find the same
answer eventually.
The theorem is complicated somewhat by the fact that the first answer that
the meta-interpreter arrives at might not be the desired one due to the
order of the clauses in the program.
To deal with this problem, we use the operator $\qryP$, which is like $\qry$,
but allows a different permutation of the program in every step.
\begin{Theorem}[Completeness]
  For any goal $\leftarrow A_1,\ldots,A_n$, if it has solution $\theta$, 
  then
  \[\begin{array}{l}
  \qryP \mathit{eval}([A_1,\ldots,A_n],\alpha,\mathit{alt}(\beta,(B_1,\ldots,B_m)),P,R).\\
  P = \mathit{success}(\gamma,(C_1,\ldots,C_k)),R = \alpha\theta.
  \end{array}\]
\end{Theorem}

Together, these two theorems show that our meta-interpreter is a conservative
extension of the conventional Prolog semantics.

\section{Related Work}

We briefly discuss the main areas of related work.

\paragraph{Continuations in $\lambda$-Prolog}

Perhaps most closely related to our work is that of
\cite{DBLP:conf/iclp/BrissetR93}, who present a continuation-passing style
semantics for $\lambda$-Prolog. Their semantics distinguishes three different
continuations: the classic success and failure continuations, and a third ``cut
failure'' continuation which cut uses to overwrite the failure continuation
with. They also expose the first two continuations, through the well-known
call-with-current-continuation operator from functional programming for the
success continuation and an analog operator for the failure continuation. They
illustrate how the latter can be used to make cut work appropriately for
meta-calls. 

A syntactic difference with our work is that they provide two seperate
operators to capture the success and failure continuation rather than a single
one. More importantly, their operators capture the full continuation while
ours capture delimited continuations. \cite{filinski} has shown that the latter 
are more expressive than the former.

\paragraph{Conjunctive Delimited Control}
Disjunctive delimited control is the culmination of a line of research on
mechanisms to modify Prolog's control flow and search, which started with the
hook-based approach of \textsc{Tor}~(\cite{DBLP:journals/scp/SchrijversDTD14})
and was followed by the development of conjunctive delimited control
for Prolog~(\cite{iclp2013,DBLP:conf/ppdp/SchrijversWDD14}).

The listing of Figure~\ref{fig:conj_reset} shows that disjunctive delimited control entirely
subsumes conjunctive delimited control. It encodes
the conjunctive reset \texttt{conj\_reset/3} in terms of our disjunctive \texttt{reset/3}, while
using the same \texttt{shift/1}. 
The conjunctive behaviour is recovered by disjoining the captured disjunctive
branch. We believe that \textsc{Tor} is similarly superseded.

\begin{figure}[t]
\begin{Verbatim}[frame=single]
  conj_reset(Goal,Ball,Cont) :-
    copy_term(Goal,GoalCopy),
    reset(GoalCopy,GoalCopy,R),
    ( R = failure -> fail
    ; R = success(BranchPattern,Branch) ->
      ( Goal = GoalCopy, Cont = 0
      ; Goal = BranchPattern, conj_reset(Branch,Ball,Cont))
    ; R = shift(X,C,BranchPattern,Branch) ->
      ( Goal = GoalCopy, Ball = X, Cont = C
      ; Goal = BranchPattern, conj_reset(Branch,Ball,Cont))
    ).
\end{Verbatim}
\caption{Encoding of conjunctive delimited control}\label{fig:conj_reset}
\end{figure}

\cite{DBLP:journals/corr/abs-1708-07081} presents a higher-level
interface for (conjunctive) delimited control on top of that of 
\cite{iclp2013}. In particular, it features \emph{prompts}, first
conceived in a Haskell implementation by \cite{prompts},
which allow shifts to dynamically specify up to what reset to capture the
continuation. We believe that it is not difficult to add a similar prompt
mechanism on top of our disjunctive version of delimited control.

\paragraph{Interoperable Engines}
\cite{DBLP:conf/padl/TarauM09}'s Interoperable Engines~
propose \emph{engines} as a means for co-operative coroutines in Prolog.  An
engine is an independent instance of a Prolog interpreter that provides answers
to the main interpreter on request.


The predicate \texttt{new\_engine(Pattern,Goal,Interactor)} creates a new
engine with answer pattern \texttt{Pattern} that will execute \texttt{Goal} and
is identified by \texttt{Interactor}.
The predicate \texttt{get(Interactor,Answer)} has an engine execute its goal
until it produces an answer (either by proving the \texttt{Goal}, or explicitly
with \texttt{return/1}).  After this predicate returns, more answers can be
requested, by calling \texttt{get/2} again with the same engine identifier.
The full interface also allows bi-directional communication between engines, but
that is out of scope here.

\begin{figure}[tb!]
\begin{Verbatim}[frame=single]
  get(Interactor,Answer) :-
    get_engine(Interactor,Engine),        % get engine state
    run_engine(Engine,NewEngine,Answer),  % run up to the next answer
    update_engine(Interactor,NewEngine).  % store the new engine state

  return(X) :- shift(return(X)).

  run_engine(engine(Pattern,Goal),NewEngine,Answer) :-
    reset(Pattern,Goal,Result),
    run_engine_result(Pattern,NewEngine,Answer,Result).

  run_engine_result(Pattern,NewEngine,Answer,failure) :-
    NewEngine = engine(Pattern,fail),
    Answer    = no.
  run_engine_result(Pattern,NewEngine,Answer,success(BPattern,B)) :-
    NewEngine = engine(BPattern,B),
    Answer    = the(Pattern).
  run_engine_result(Pattern,NewEngine,Answer,S) :-
    S = shift(return(X),C,BPattern,B)
    BPattern  = Pattern,
    NewEngine = engine(Pattern,(C;B)),
    Answer    = the(X).
\end{Verbatim}
\caption{Interoperable Engines in terms of delimited control.}
\label{fig:interop-engines}
\end{figure}

Figure~\ref{fig:interop-engines} shows that we can implement the \texttt{get/2}
engine interface in terms of delimited control (the full code is available
in the online repository).
The opposite, implementing disjunctive delimited control with engines, seems
impossible as engines do not provide explicit control over the disjunctive
continuation. Indeed, \texttt{get/2} can only follow Prolog's natural
left-to-right control flow and thus we cannot, e.g., run the disjunctive
continuation before the conjunctive continuation, which is trivial with
disjunctive delimited control.

\paragraph{Functional Programming Models of Nondeterminism and Backtracking}

Prolog-style nondeterminism, and in particular its backtracking approach, have
been widely studied from a Functional Programming perspective and various
abstraction mechanisms have been proposed to capture it.

\cite{DBLP:journals/ngc/Carlsson84} has shown how to implement Prolog-style backtracking with a single
``success'' continuation.  A decade later,
\cite{DBLP:journals/toplas/Gudeman92} uses a second ``failure'' continuation in
the denotational semantics of the goal-oriented Icon language to conveniently
express control flow manipulations.

\cite{DBLP:conf/fpca/Wadler85} has shown how to encode backtracking with lazy
lists, \cite{DBLP:journals/scp/Spivey90} noticed that this fit the category
theoretical structure of monads which was further expanded upon by
\cite{DBLP:conf/lfp/Wadler90}. Later, \cite{DBLP:conf/mpc/Hinze12} has shown that the lazy list monad
and the two-continuation approach, which also has monadic structure, are 
two equivalent representations obtained from the same adjunction.

Another related development is that of \emph{algebraic effects \& handlers},
pioneered by \cite{DBLP:journals/corr/PlotkinP13}, a high-level mechanism for
modelling side effects such as nondeterminism. \cite{DBLP:conf/icfp/KammarLO13} have shown that this
mechanism can be implemented both in terms of delimited control and of the
so-called free monad. The latter reifies the computation as a tree-like data
structure. We can find precursors of this approach in Curry's encapsulated
search tree (\cite{DBLP:journals/jflp/BrasselHH04}) and the monadic constraint
programming framework (\cite{DBLP:journals/jfp/SchrijversSW09}), which both
expose an explicit search tree that can be manipulated to obtain various search
strategies.

\paragraph{Tabling without non-bactrackable variables}
Tabling~(\cite{swift2012,yap}) is a well-known technique that eliminates the
sensitivity of SLD-resolution to clause and goal ordering, allowing a larger
class of programs to terminate.
As a bonus, it may improve the run-time performance (at the expense of
increased memory consumption).

One way to implement tabling---with minimal engineering impact to the Prolog
engine---is the tabling-as-a-library approach proposed by 
\cite{iclp2015}.
This approach requires (global) mutable variables that are not erased by
backtracking to store their data structures in a persistent manner.
With the new \texttt{reset/3} predicate, this is no longer needed, as
(non-backtracking) state can be implemented in directly with disjunctive
delimited control.

\paragraph{Probabilistic Logic Programming}
The implementation techniques used by existing probabilistic logic programming
systems use more elaborate and sophisticated approaches than the lightweight technique we have presented.
PRISM is implemented on top of B-Prolog (\cite{DBLP:journals/jair/SatoK01}) and uses its tabling
mechanism to execute programs that are transformed to collect rather than
execute \texttt{msw/2} calls. Based on the answer tables, support graphs are 
constructed from which the probabilities are computed.

The first version of ProbLog (\cite{DBLP:journals/tplp/KimmigDRCR11}) used a
similar tabling-based approach to collect all the proofs of a goal and
post-process these. The present version of ProbLog (\cite{DBLP:journals/tplp/FierensBRSGTJR15}) converts a program to a
weighted boolean formula, then converts this to a circuit in deterministic,
decomposable negation normal form (\cite{DBLP:conf/ecai/Darwiche04}) which can be directly evaluated with
structural recursion.

\section{Conclusion and Future Work}
We have presented \emph{disjunctive delimited control}, an extension to
delimited control that takes Prolog's non-deterministic nature into account.
This is a conservative extension that enables implementing disjunction-related
language features and extensions as a library.

In future work, we plan to explore a WAM-level implementation of disjunctive
delimited control, inspired by the stack freezing functionality of tabling
engines, to gain access to the disjunctive continuations efficiently.
Similarly, the use of \texttt{copy\_term/2} necessitated by the current
API has a detrimental impact on performance, which might be overcome by a
sharing or shallow copying scheme.

Inspired by the impact of conjunctive delimited control, which has brought
tabling to SWI-Prolog, we believe that further development of disjunctive
delimited control is worthwhile. Indeed, it has the potential of bringing
powerful disjunctive control abstractions like branch-and-bound search to a
wider range of Prolog systems. 



\paragraph{Acknowledgments}
We are grateful to Paul Tarau and the anonymous LOPSTR 2021 reviewers for their
helpful feedback.
Part of this work was funded by FWO grant G0D1419N and by KU Leuven grant C14/20/079.

\bibliographystyle{tlplike}
\bibliography{disj}

\label{lastpage}
\end{document}